\documentclass[twocolumn,showpacs,preprintnumbers,prb,superscriptaddress]{revtex4}

\usepackage{graphicx}

\begin{document}
\title{Electron correlation and two dimensionality in the spin-density-wave phase
of (TMTTF)$_2$Br under pressure}
\author{A. Ishikawa}
 \email{a_ishikawa@nucc.cc.nagoya-u.ac.jp}
 \altaffiliation[Present address: ]{Venture Business Laboratory, 
 Nagoya University, Furo-cho, Chikusa-ku, Nagoya 464-8601, Japan}
\author{N. Matsunaga}
 \email{mat@phys.sci.hokudai.ac.jp}
 \homepage{http://phys.sci.hokudai.ac.jp/LABS/nomura/english.html}
\author{K. Nomura}
\affiliation{Division of Physics, Hokkaido University, Sapporo 060-0810, Japan}
\author{T. Sasaki}
\affiliation{IMR, Tohoku University, Sendai 980-8577, Japan}
\author{T. Nakamura}
\affiliation{Institute for Molecular Science, Myodaiji, Okazaki 444-8585, Japan}
\author{T. Takahashi}
\affiliation{Department of Physics, Gakushuin University, Mejiro 1-5-1, Toshima-ku, 
Tokyo 171-8588, Japan}
\author{G. Saito}
\affiliation{Division of Chemistry, Kyoto University, Sakyo-ku, Kyoto 606-8502, 
Japan}
\date{Submitted April 28, 2003}

\begin{abstract}
The incommensurate spin-density-wave (SDW) phase in (TMTTF)$_2$Br was investigated 
through transport measurements under pressure and magnetic fields parallel to the  $c^{\ast}$ axis. 
For the incommensurate SDW phase of (TMTTF)$_2$Br stabilized above 0.5 GPa,
the SDW transition temperature $T_{\rm SDW}$ increases with the applied magnetic field. 
The field dependence of $T_{\rm SDW}$ is described by a quadratic behavior 
and the coefficient of the quadratic term increases with increasing pressure. 
These results are consistent with the prediction of the mean-field theory based 
on the suppression of the SDW transition by two-dimensionality.    
From the relation between 
the coefficient of the quadratic term and $T_{\rm SDW}$ at zero magnetic field, 
we determined the role of electron correlation and two dimensionality in the SDW phase
of (TMTTF)$_2$Br under pressure and found that
the SDW transition in (TMTTF)$_2$Br can be well explained within the mean field theory 
by taking into account the reduction of the coupling constant $N(0)I$ by pressure.

\end{abstract}

\maketitle

The family of organic salts (TMT$C$F)$_{2}X$  
($C$ = Se or S and $X$ = PF$_{6}$, AsF$_{6}$, ClO$_{4}$, Br, etc.) 
show quasi-one-dimensional (Q1D) electric properties and 
have rich ground states, 
spin-Peierls, antiferromagnetism (AF), spin density wave (SDW), 
superconductivity, etc.
The ground state in these salts is influenced 
by the different donors and anions that constitute the compounds. 
A universal phase diagram for the TMT$C$F salts as a function of pressure 
has been proposed by J\'{e}rome.~\cite{jerome} 
The differences of the donors and the anions 
change the distance between molecules, transfer integral, etc.\
and this means that these act as a chemical pressure
in the universal phase diagram. 
With increasing pressure, the electron transfer along the interchain direction 
increases, and the two dimensionality of the system increases.
In addition to the role of the two dimensionality, it is also well known that the electron correlation must be strong 
to cause such a rich phase diagram in the Q1D compounds. 

The ground state of the sulfur-based salt $\rm (TMTTF)_{2}Br$ is
antiferromagnetism (AF) at ambient pressure.~\cite{barthel}
In the temperature range above this AF phase, 
the system shows a broad minimum of resistivity $\rho_{\rm {min}}$ at about $T_\rho$ = 100 K. 
Between this temperature and the AF transition temperature, the system has a charge-localized (CL) state, 
in which the resistance increases with decreasing temperature.
On applying the pressure, the ground state changes from the AF state to the 
incommensurate SDW state above about 0.5 GPa and $\rho_{\rm {min}}$ vanishes. 
It is known that this incommensurate SDW state in (TMTTF)$_2$Br is essentially the same as that in (TMTSF)$_2X$,
which is stabilized by the nesting of the Fermi surface in the Q1D electron band. 

When the magnetic field is applied to the SDW state suppressed by the imperfect nesting of Fermi surface, 
it has been predicted that $T_{\rm {SDW}}$ increases nearly quadratically with the field in low magnetic fields
and shows a saturation behavior to the transition temperature for the perfect nesting case $T_{\rm {SDW_0}}$
in high magnetic fields.~\cite{montambaux}
In fact, the quadratic magnetic field dependence and the saturation behavior have been confirmed by the experiments for 
(TMTSF)$_2$PF$_6$ (Refs.\onlinecite{Mat-PF6,Mat-PF6-2}) and quenched (TMTSF)$_2$ClO$_4$
(Refs.\onlinecite{Mat-ClO4,Mat-ClO4-2,Qualls}).
In a previous paper, however, we have estimated $T_{\rm {SDW_0}}$ = 16 K
for $\rm (TMTSF)_{2}PF_{6}$ under low pressure.~\cite{Mat-PF6}
This value of $T_{\rm {SDW_0}}$ is slightly small to explain the universal phase diagram
using the simplest model of the mean-field (MF) theory.

In this paper, we present the results of resistivity measurements for the incommensurate SDW phase 
in (TMTTF)$_2$Br under pressure and magnetic field, and discuss the validity of the universal phase diagram
for the TMT$C$F salts through the role of electron correlation 
and two dimensionality in the SDW phase of (TMTTF)$_2$Br under pressure.


Single crystals of $\rm (TMTTF)_{2}Br$ were synthesized by the electrochemical method. 
The resistance measurements were performed by a dc four-wire method 
along the highly conducting $a$ axis. 
Electric leads of 10 $\mu$m gold wire were attached with silver paint onto gold evaporated 
contacts.
The current contacts covered the whole areas of both ends of the sample for a uniform current.
The typical sample size was 0.4 $\times$ 0.2 $\times$ 0.02 mm$^3$, 
where the length along the $a$-axis direction is that between voltage contacts. 
The temperature was measured using a Cernox CX-1050-SD
resistance thermometer calibrated by a capacitance sensor in magnetic fields.
The pressure was applied using the beryllium-copper clamp cell up to 2.1 GPa. 
We used Daphne 7373 oil as a pressure medium. 
It is known that the pressure decreases 
at low temperature by about 0.15 GPa from any initial pressure at room temperature 
due to the solidification of this medium.~\cite{Murata} 
The magnetic field was applied along the $c^\ast$ axis up to 24 T
in a hybrid magnet at the High Magnetic Field Laboratory, IMR, Tohoku University.


\begin{figure}
\includegraphics[trim=65mm 110mm 65mm 110mm, width=3.0in]{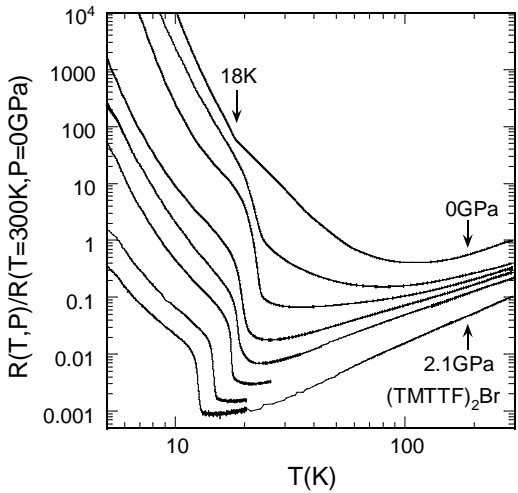}
\caption{Temperature dependence of the resistance normalized by the resistance 
at ambient pressure and room temperature. The pressure is from top to bottom, 
0, 0.2, 0.3, 0.5, 0.75, 1.0, 1.65, 2.1 GPa, respectively. 
} \label{fig:R-norm}
\end{figure}

\begin{figure}
\includegraphics[trim=15mm 85mm 24mm 41mm, width=3.0in]{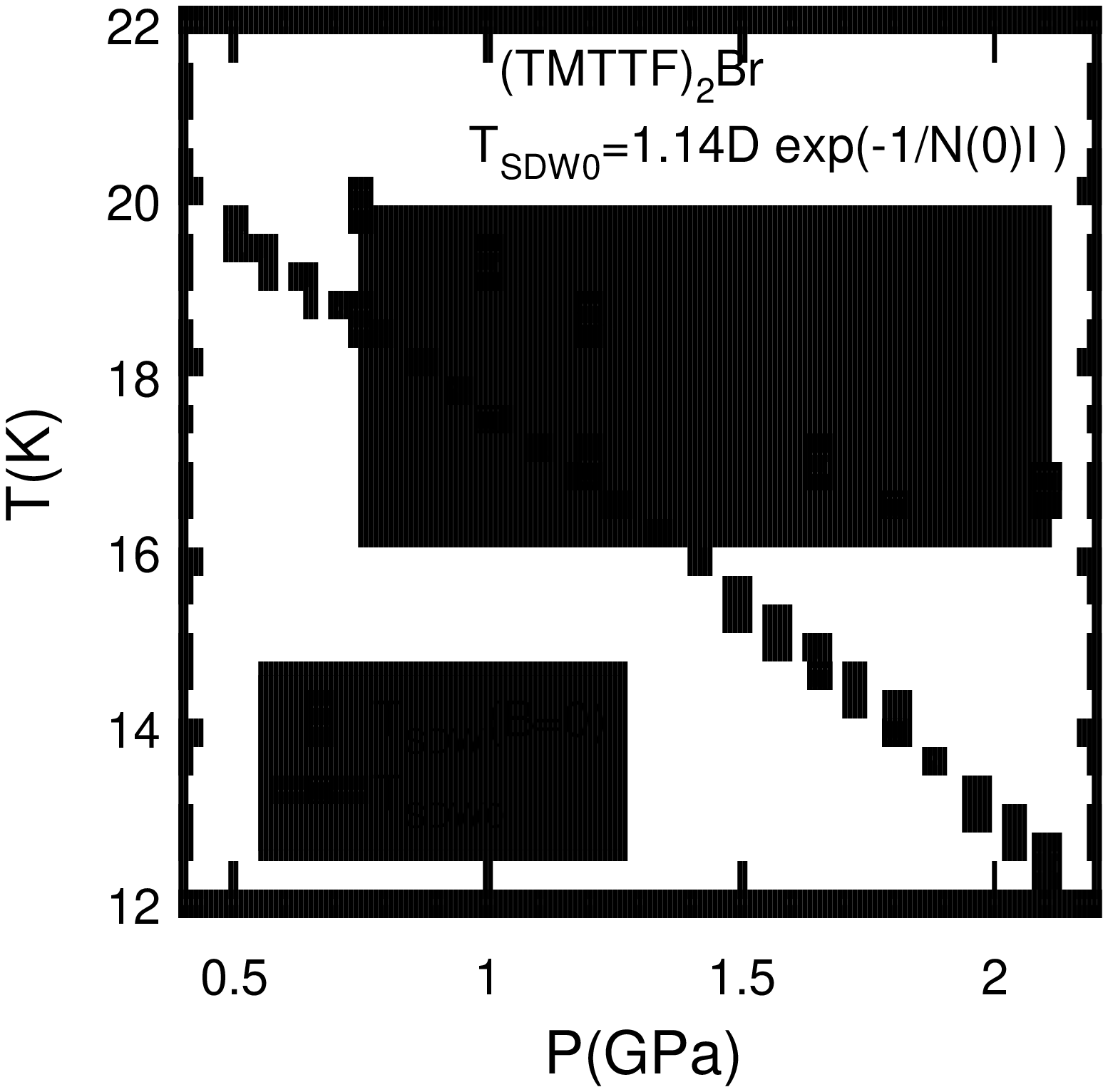}
\caption{Pressure dependence of the SDW transition temperature at zero magnetic field $T_{\rm SDW}$(0)
and the SDW transition temperature for the perfect nesting case $T_{\rm {SDW_0}}$ 
estimated by fitting using the mean-field theory
taking into account the magnetic-field dependence of the SDW transition temperature.
The solid line is the best fit of the linear function.
The dashed line is guide to the eye.
} \label{fig:P-Tsdw0}
\end{figure}

Figure \ref{fig:R-norm} shows the temperature dependence of the resistance along the $a$ axis in 
$\rm (TMTTF)_{2}Br$ for various pressures. 
At ambient pressure, the resistance shows the broad minimum at about 100 K, 
corresponds to the crossover from the metallic to the CL state. 
Then, the resistance increases with decreasing temperature 
and shows a kink at 18 K. 
From the NMR measurements \cite{barthel}, it is well known that 
$\rm (TMTTF)_{2}Br$ exhibits an AF transition at 14 K. 
However, the kink temperature is 4 K higher than this
and $d({\rm ln}R)/d(1/T)$ below 18 K is almost flat.
The resistance shows no characteristic structure at 14 K. 
This indicates that the AF ordering does not change 
the behavior of conductivity. 
The kink structure at 18 K in the resistance at ambient pressure may be 
caused by charge disproportionation.~\cite{Ishikawa}
The temperature for the resistance minimum $T_{\rm min}$ rapidly decreases with increasing pressure up to 0.3 GPa 
and vanishes at about 0.4 GPa. 
These results are consistent with a previous report.~\cite{klemme1} 
As shown in Fig.\ \ref{fig:R-norm}, the kink at 18 K in the resistance at ambient pressure changes to the broad peak 
structure in the derivative of the logarithm of the resistance with $1/T$. 
The temperature of this peak in $d({\rm ln}R)/d(1/T)$ increases with increasing 
pressure and shows a maximum of 23 K at 0.3 GPa. 
Above 0.3 GPa, the peak temperature decreases from 23 K at 0.3 GPa to 19.5 K at 0.5 GPa.
A detailed discussion of this phase diagram below 0.5 GPa is contained in a previous paper.~\cite{Ishikawa} 

Above 0.5 GPa the peak temperature still decreases with pressure, 
but the rate of decrease becomes smaller. 
It has been confirmed by the NMR measurements \cite{hisano} that the SDW with the incommensurate wave vector 
is stabilized above 0.5 GPa.
With decreasing temperature, the resistance 
shows a steep increase associated with the SDW transition and thermally 
activated behavior at low temperature as shown in Fig.\ \ref{fig:R-norm}.
The SDW transition temperatures in zero field $T_{\rm {SDW}}$(0) determined 
from the peak of the derivative of the logarithm of 
the resistance with $1/T$ $d({\rm ln}R)/d(1/T)$ are plotted against field in Fig.\ \ref{fig:P-Tsdw0}.
With increasing pressure, $T_{\rm SDW}$(0) decreases from 19.5 K at 0.5 GPa to 12.5 K at 2.1 GPa.
This pressure dependence of $T_{\rm {SDW}}$(0) above 0.5 GPa is qualitatively consistent with 
the prediction of the MF theory based on the imperfect nesting of the Fermi surface.

\begin{figure}
\includegraphics[trim=0mm 47mm 19mm 5mm, width=3.0in]{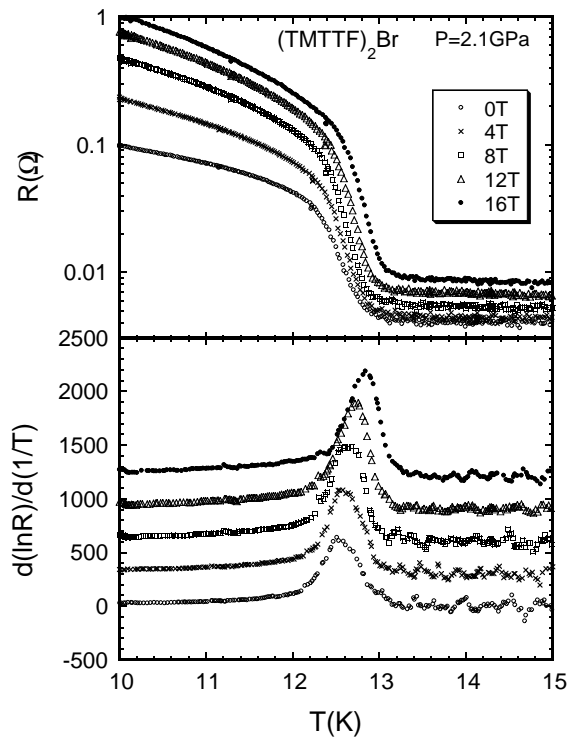}
\caption{Top: Temperature dependence of the resistance in 
$\rm (TMTTF)_{2}Br$ for various magnetic fields at 2.1 GPa. 
Bottom: The derivative of the logarithm of the resistance by $1/T$
offset from zero for clarity.
} \label{TD2-1}
\end{figure}

\begin{figure}
\includegraphics[trim=7mm 60mm 23mm 60mm, width=3.0in]{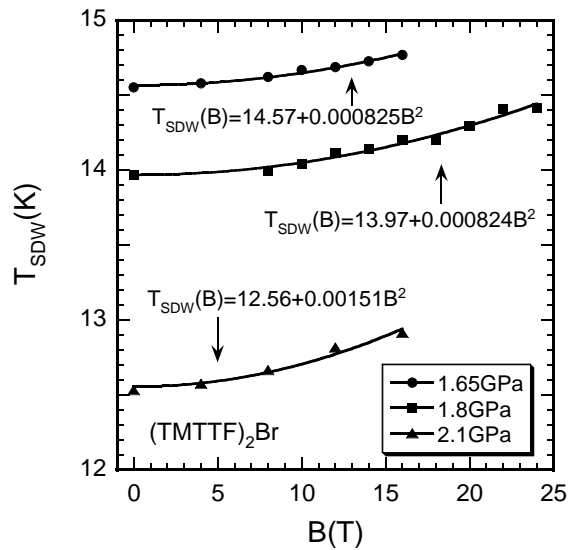}
\caption{Magnetic-field dependence of the SDW transition temperature $T_{\rm SDW}$ for various pressures.
The solid lines are the best fit of the quadratic function.
} \label{MFD}
\end{figure}

Figure \ref{TD2-1} shows the temperature dependence of the resistance of 
$\rm (TMTTF)_{2}Br$ along the $a$-axis under various magnetic fields at 2.1 GPa. 
In each field, the resistance shows the increase with decreasing temperature associated
with the SDW gap. 
The SDW transition temperature $T_{\rm SDW}$ in a magnetic field, determined from 
$d({\rm ln}R)/d(1/T)$ (the bottom of Fig.\ \ref{TD2-1}), is shown in Fig.\ \ref{MFD}. 
We find that, with increasing magnetic field along the $c^\ast$ axis, $T_{\rm SDW}$ increases
at various pressures.
The magnetic-field dependence of $T_{\rm SDW}$ is described by the quadratic 
function:
$T_{\rm SDW}(B)=T_{\rm SDW}(0)+CB^2$,
where $C$ is constant, for each pressure.
The coefficient of the quadratic term $C$ increases with increasing pressure. 
All the results between 0.75 GPa and 2.1 GPa, are well fitted by the quadratic function without any 
saturation behavior even at 24 T. 
Such a behavior is also found in TMTSF salts \cite{Mat-PF6,Mat-ClO4} and 
is consistent with the prediction of the MF theory.~\cite{montambaux}
As a result, we conclude that the SDW transition in $\rm (TMTTF)_{2}Br$ is an ordinary SDW transition based on 
the nesting of the Fermi surface as for (TMTSF)$_{2}X$. 

\begin{figure}
\includegraphics[trim=10mm 98mm 20mm 30mm, width=3.0in]{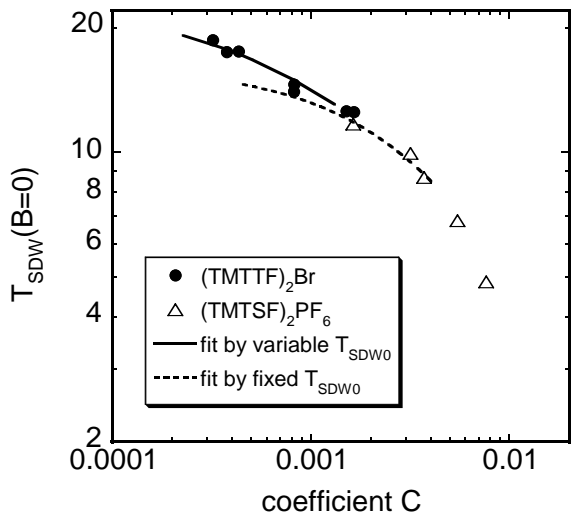}
\caption{The relation between the SDW transition temperature at zero magnetic field $T_{\rm SDW}$(0) and 
the coefficient $C$ of quadratic term in $T_{\rm SDW}$ in (TMTTF)$_2$Br and 
(TMTSF)$_2$PF$_6$. The broken line represents the theoretical curve with $T_{\rm SDW0}$ = 16 K. 
The solid line represents the theoretical curve taking into account the 
pressure dependence of $N(0)I$ shown in Fig.\ \ref{fig:P-Tsdw0}. 
} \label{TsdwvsC}
\end{figure}

Figure \ref{TsdwvsC} shows the relation between the coefficient of the quadratic term $C$ and 
$T_{\rm SDW}$ at zero magnetic field $T_{\rm SDW}$(0) for $\rm (TMTTF)_{2}Br$. 
The coefficient $C$ is mainly determined by the two dimensionality of the electron band in the MF theory, 
however, $C$ also depends on other parameters, e.g., Fermi velocity $v_{\rm F}$. 
In the figure, the result for (TMTSF)$_2$PF$_6$ is also shown for comparison.~\cite{Mat-PF6}
The results of $\rm (TMTTF)_{2}Br$ under high pressure seem 
to smoothly connected to those of $\rm (TMTSF)_{2}PF_{6}$, and the SDW-superconductivity 
phase diagram in $\rm (TMTTF)_{2}Br$ above 2.1 GPa\cite{Creuzet,Balicas} is quite similar to
that in $\rm (TMTSF)_{2}PF_{6}$ under pressure.~\cite{Jerome}
Therefore, we expect that the parameters in $\rm (TMTTF)_{2}Br$ at 2.1 GPa to be close to those 
in $\rm (TMTSF)_{2}PF_{6}$ at ambient pressure. 
From the observed curvature of the line for the relation between $T_{\rm {SDW}}$(0) 
and $C$ for (TMTSF)$_2$PF$_6$, 
we had determined several parameters for the SDW transition in the MF theory, \cite{maki}
assuming that the SDW transition temperature for the perfect nesting case $T_{\rm SDW_0}$ 
is independent of pressure as in a previous paper.~\cite{Mat-PF6} 
The dashed line in Fig.\ \ref{TsdwvsC} is the fit by this theory to the experimental data of $\rm (TMTSF)_{2}PF_{6}$ and the agreement is good.
The fit gives  $T_{\rm {SDW_0}}$ = 16 K and $v_{\rm F}$ = 1.03 $\times$ 10$^5$ m/s for $\rm (TMTSF)_{2}PF_{6}$,
 with a lattice parameter along the $b^{\prime}$ axis of 7.7 $\times$ 10$^{-10}$ m.
However, the data for $\rm (TMTTF)_{2}Br$ in the region where  $T_{\rm {SDW}}$(0) is higher than 13 K,
 corresponding to 1.7 GPa, are slightly above the calculated dashed curve.
This fact indicates that all the data for $\rm (TMTTF)_{2}Br$ and 
$\rm (TMTSF)_{2}PF_{6}$ cannot be described by the constant value of $T_{\rm SDW_0}$ with the pressure. 
Even if we consider only the results for $\rm (TMTTF)_{2}Br$, 
the almost linear relation between $T_{\rm {SDW}}$(0) and $C$ in a log-log scale, shown in Fig.\ \ref{TsdwvsC},
 is also hardly explained by the MF theory with constant $T_{\rm {SDW_0}}$. 
In such a simple model, it is expected that $T_{\rm {SDW}}$(0) is almost 
 constant for the imperfect nesting parameter $\epsilon_0$/$\Delta_0$ 
 in the region where $\epsilon_0$/$\Delta_0$ is small and $T_{\rm {SDW}}$(0) is close to $T_{\rm {SDW_0}}$,
 in contrast to the observed behavior. 
As a result, the observed slightly steep and straight dependence of $T_{\rm {SDW}}$ suggests 
that $T_{\rm {SDW_0}}$ varies with the pressure. 

From the BCS relation in the MF theory, $T_{\rm SDW_0}$ is given as 
$k_{\rm B}T_{\rm SDW_0}=1.14D\,{\rm exp}(-1/N(0)I)$, 
where $D$, $N(0)$, and $I$ are the bandwidth, the density of state 
at Fermi level, and the on-site Coulomb energy, respectively. 
With increasing pressure, we can expect that 
(a) the on-site Coulomb energy $I$ decreases 
due to the increase of the screening of Coulomb potential, 
(b) the density of state $N(0)$ decreases due to the increase of the transfer energy, 
and (c) the bandwidth $D$ increases due to the increase of the transfer energy along the $a$ axis.
The contribution of $N(0)I$ to $T_{\rm SDW_0}$ is expected to be larger than that of $D$ 
because the variation of $N(0)I$ affects $T_{\rm SDW_0}$ exponentially.
After all, we can expect $T_{\rm SDW_0}$ to decrease due to the decrease of the coupling 
constant $N(0)I$ with increasing pressure.
Indeed, the reduction of $N(0)I$ by applying pressure was suggested 
from the NMR measurements in the metallic phase of (TMT$C$F)$_2X$ under the pressure.~\cite{wzietek}
It is natural to consider that $T_{\rm {SDW_0}}$ decreases as the applied pressure becomes larger, 
although a quantitative estimate is difficult at present. 
If such a situation is realized in the present system, 
the observed upward deviation of data and its fairly steep dependence in the region,
where $T_{\rm {SDW}}$(0) is higher than 13 K, can be explained.
  
Assuming $v_{\rm F}=1.03\times 10^5{\rm m/s}$ independent of 
the pressure,~\cite{Comment1} $T_{\rm SDW_0}$ at each pressure can be estimated by fitting with the MF theory. 
Figure \ref{fig:P-Tsdw0} also shows the pressure dependence of
$T_{\rm SDW_0}$ obtained in this way. 
The value of $T_{\rm SDW_0}$ decreases 
with increasing pressure, from 20 K at 0.75 GPa to 16.8 K 
at 2.1 GPa, and the decrease of $T_{\rm SDW_0}$ is approximately linear with the
increase of pressure as denoted by the solid line in Fig.\ \ref{fig:P-Tsdw0}. 
Using this linear dependence for $T_{\rm SDW_0}$, the magnetic-field dependence of $T_{\rm SDW}$ in $\rm (TMTTF)_2Br$ 
is well explained by the solid line in Fig.\ \ref{TsdwvsC}. 
With increasing pressure, it is expected that
$v_{\rm F}$ increases due to the increase of the transfer energy along the $a$ axis.
In this analysis, we assume that $v_{\rm F}$ is constant against the pressure;
it is expected that $v_{\rm F}$ does not show any large pressure dependence in the present conditions
and a small pressure dependence of $v_{\rm F}$ does not change our conclusion.~\cite{Comment2} 
As a result, the magnetic-field dependence of $T_{\rm SDW}$ in $\rm (TMTTF)_2Br$ 
is consistent with the prediction of the MF theory by 
taking into account the decrease of $T_{\rm SDW_0}$ due to the reduction of $N(0)I$ by pressure. 

In the case of (TMTSF)$_2$PF$_6$, the pressure dependence of $T_{\rm {SDW}}$(0)
is mainly determined by the imperfect nesting parameter $\epsilon_0$/$\Delta_0$, 
because $\epsilon_0$/$\Delta_0$ is fairy large and $T_{\rm {SDW}}$(0) 
shows a rapid decrease with increasing $\epsilon_0$/$\Delta_0$.
In such a situation, the reduction of $N(0)I$ with pressure makes only a small contribution to $T_{\rm {SDW}}$(0). 
As a result, the simplest model with constant $T_{\rm {SDW_0}}$
reproduces the observed relation between $T_{\rm {SDW}}$(0) and $C$  well for (TMTSF)$_2$PF$_6$,
as described in a previous paper.~\cite{Mat-PF6}
Accordingly, as shown in Fig.\ \ref{TsdwvsC}, the incommensurate SDW phase
in both (TMTSF)$_2$PF$_6$ and (TMTTF)$_2$Br is understood systematically and quantitatively in the common pressure axis, 
where its origin corresponding to ambient pressure is shifted for these two salts, 
with the MF theory based on the nesting of the Fermi surface taking into account 
the reduction of the coupling constant with pressure.

In summary, resistivity measurements have been performed under pressure and in magnetic 
fields in $\rm (TMTTF)_{2}Br$. 
Above 0.5 GPa, $\rm (TMTTF)_{2}Br$ exhibits an incommensurate SDW state at low temperatures. 
With increasing pressure, 
the SDW transition temperature $T_{\rm SDW}$ decreases from 19.5 K at 0.5 GPa to 
12.5 K at 2.1 GPa. 
When the magnetic field is applied to this SDW phase, substantially suppressed by the pressure,
$T_{\rm SDW}$ increases quadratically with the field and the coefficient $C$ of the quadratic term 
increases with increasing pressure. 
These results are consistent with the prediction of the MF 
theory based on the nesting of the Fermi surface in the Q1D 
electronic band. 
From the analysis of the relation between 
the coefficient $C$ and $T_{\rm SDW}$ at zero magnetic field, 
the magnetic field and pressure dependence of $T_{\rm SDW}$ in 
$\rm (TMTTF)_{2}Br$ can be well explained by the MF theory by 
taking into account the reduction of the coupling constant $N(0)I$ with pressure. 
As a result, the incommensurate SDW phase 
in both (TMTSF)$_2$PF$_6$ and (TMTTF)$_2$Br is well described by the MF theory 
taking into account electron correlation and two dimensionality of the system in the universal phase diagram.

We would like to thank Professor K. Murata for technical advice on the high-pressure measurement.


\begin{thebibliography}{99}
\bibitem{Comment1}
With the increasing pressure, we can expect the increase of the transfer energy along the $a$ axis, 
which works as the increase of $v_{\rm F}$, and the decrease of the lattice parameter along the $a$ axis, 
which works as the decrease of $v_{\rm F}$.
Although a quantitative estimation of the pressure dependence of $v_{\rm F}$ is difficult at present,
a small pressure dependence does not change our conclusion.~\cite{Comment2}
\bibitem{Comment2}
If the value of $v_{\rm F}$ increases 20\% with increasing pressure from 0.75 GPa to  2.1 GPa,
$T_{\rm SDW_0}$ at 0.75 GPa increases from 20.0 K to 20.6 K.
\end{thebibliography}
\end{document}